# 4D Specialty Approximation: Ability to Distinguish between Related Specialties

Nadine Rons*

* Nadine.Rons@vub.ac.be
Research Coordination Unit, Vrije Universiteit Brussel (VUB), Pleinlaan 2, B-1050 Brussels (Belgium)

## ABSTRACT
Publication and citation patterns can vary significantly between related disciplines or more narrow specialties, even when sharing journals. Journal-based structures are therefore not accurate enough to approximate certain specialties, neither subject categories in global citation indices, nor cell sub-structures (Rons, 2012). This paper presents first test results of a new methodology that approximates the specialty of a highly specialized seed record by combining criteria for four publication metadata-fields, thereby broadly covering conceptual components defining disciplines and scholarly communication. To offer added value compared to journal-based structures, the methodology needs to generate sufficiently distinct results for seed directories in related specialties (sharing subject categories, cells, or even sources) with significantly different characteristics. This is tested successfully for the sub-domains of theoretical and experimental particle physics. In particular analyses of specialties with characteristics deviating from those of a broader discipline embedded in can benefit from an approach discerning down to specialty level. Such specialties are potentially present in all disciplines, for instance as cases of peripheral, emerging, frontier, or strategically prioritized research areas.

## METHODOLOGY
The methodology uses four publication metadata-fields (references, authors, title, source) that are generally available in global citation indices such as Thomson Reuters' Web of Science (WoS) and Elsevier's Scopus. These are related to four of the six conceptual components defining disciplines as synthesized by Sugimoto and Weingart (2015) (cognitive, social, communicative, separatedness; the other two being: tradition, institutional) and to the four facets of the framework for bibliometric analysis of scholarly communication as proposed by Ni, Sugimoto and Cronin (2013) (artifact, producer, concept, gatekeeper). Combinations of some of these dimensions have been used previously to identify publication sets associated to particular research areas in various contexts (e.g. mapping, normalization, information retrieval). To the best of the author's knowledge, the proposed methodology is the first to bibliometrically approximate a specialty using criteria with this breath of coverage of related conceptual components. It can be applied to publication records as specialized as those of an individual scientist or a team's research programme, provided that the seed directory is enlarged with publications referred to (diversifying authors while mainly adding publications in the same specialty, or at least in case of non-interdisciplinary research).

In a first phase, most frequently occurring 'key values' are selected in each dimension (references, authors, title words, cells) until a pre-set percentage (coverage threshold) of publications in the seed directory is covered by key values. In a second phase, the specialty is approximated by the set of publications covered by key values in at least three of the four dimensions. In both phases coverage of a publication by key values in a particular dimension requires the publication to be associated to at least one key value for authors, references, and cells, and to at least two key values for title words. The possibility not to be associated to key values in one of the four dimensions prevents exclusions based on that dimension only of otherwise complying publications (false negatives). The required association to key values in at least three dimensions prevents inclusions based on one or two dimensions only of otherwise non-complying publications (false positives). The combination of dimensions also allows complexity per dimension to remain low. In the calculations for this paper, the coverage threshold was set to 80% for all dimensions, and key values were limited to words of at least five characters, to references in WoS identified via DOI, and to reprint authors (processed based on name and first initial, excluding frequently occurring names).

Figure 1. Proximity in publication venues of two team leaders in theoretical and experimental particle physics

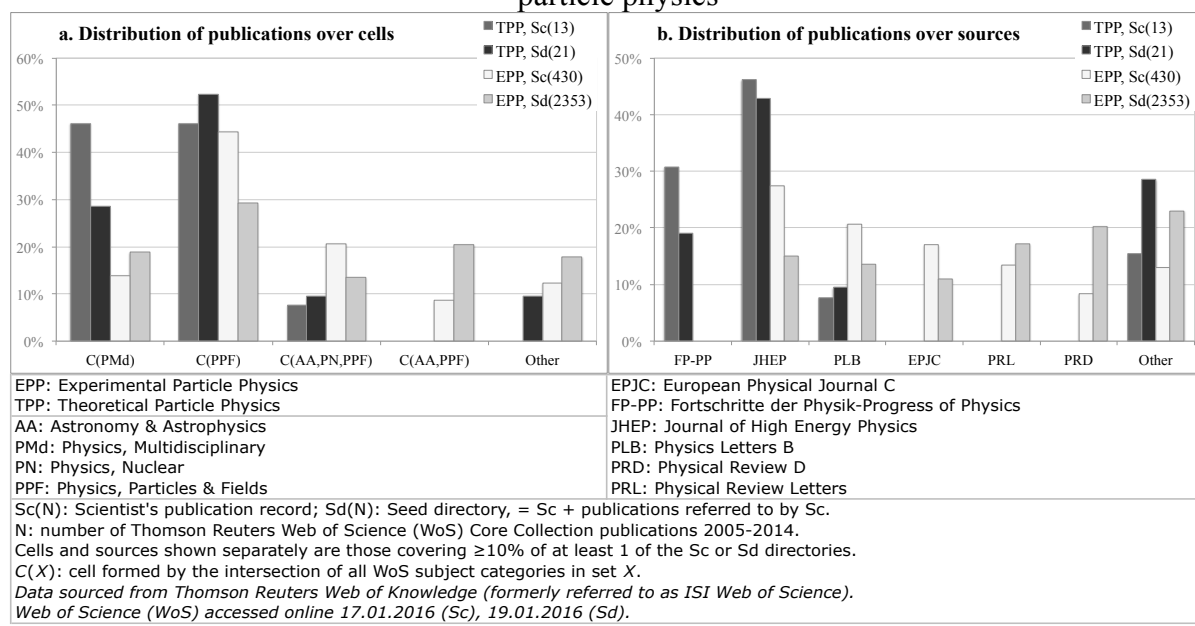

EPP: Experimental Particle Physics
TPP: Theoretical Particle Physics
AA: Astronomy & Astrophysics
PMd: Physics, Multidisciplinary
PN: Physics, Nuclear
PPF: Physics, Particles & Fields

EPJC: European Physical Journal C
FP-PP: Fortschritte der Physik-Progress of Physics
JHEP: Journal of High Energy Physics
PLB: Physics Letters B
PRD: Physical Review D
PRL: Physical Review Letters

Sc(N): Scientist's publication record; Sd(N): Seed directory, = Sc + publications referred to by Sc.
N: number of Thomson Reuters Web of Science (WoS) Core Collection publications 2005-2014.
Cells and sources shown separately are those covering ≥10% of at least 1 of the Sc or Sd directories.
$C(X)$: cell formed by the intersection of all WoS subject categories in set $X$.
Data sourced from Thomson Reuters Web of Knowledge (formerly referred to as ISI Web of Science).
Web of Science (WoS) accessed online 17.01.2016 (Sc), 19.01.2016 (Sd).

**RESULTS**
Figure 1 illustrates the proximity in publication venues of two team leaders in theoretical and experimental particle physics, substantially sharing cells and even sources published in and referred to. These sub-domains are nevertheless known to strongly differ in attained numbers of co-authors and citations, much higher for experimental particle physics. In the related subject categories, cells, and even sources, these different cultures are blended. The scientists' 4D specialty approximations keep these traditions apart (no overlap), and strongly differ in attained levels of co-authors and citations (a few co-authors and several hundred citations for theoretical particle physics, versus several thousand for experimental particle physics), reflecting the known differences between these sub-domains (Figure 2).

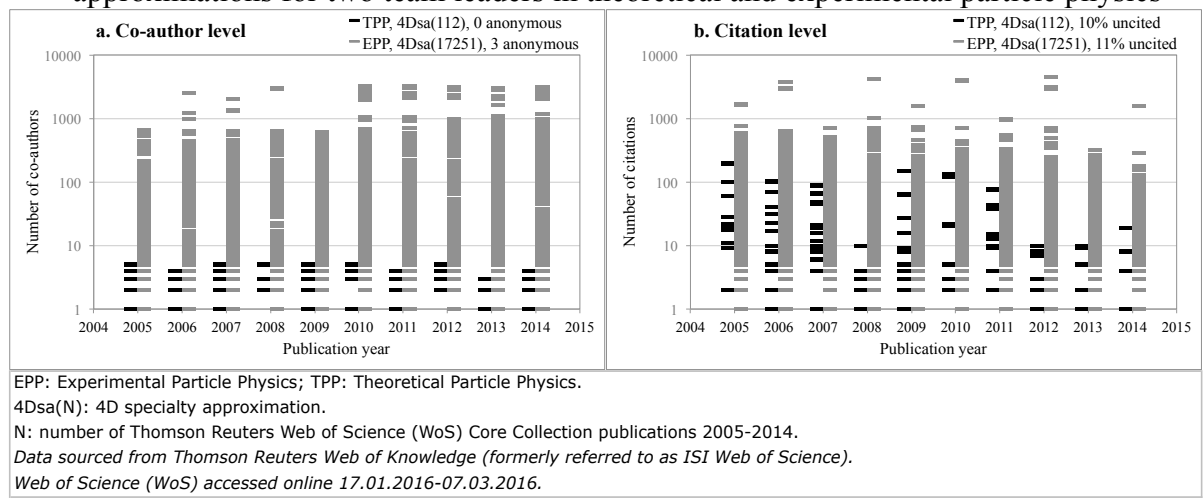

Figure 2. Differences in attained co-author and citation levels between 4D specialty approximations for two team leaders in theoretical and experimental particle physics

EPP: Experimental Particle Physics; TPP: Theoretical Particle Physics.
4Dsa(N): 4D specialty approximation.
N: number of Thomson Reuters Web of Science (WoS) Core Collection publications 2005-2014.
*Data sourced from Thomson Reuters Web of Knowledge (formerly referred to as ISI Web of Science).*
*Web of Science (WoS) accessed online 17.01.2016-07.03.2016.*

**DISCUSSION**

Specialty approximations can provide information for various analyses with quantitative aims (e.g. reference values for normalized indicators, thresholds for outstanding performance) or qualitative aims (e.g. lists of potential peers, benchmarks, literature of interest). Whether a specialty approximation is sufficiently accurate (precise in delineation and complete in coverage) depends on the information to be derived from it, and on how strongly this information varies between related specialties. Also sub-specialties (e.g. dedicated to specific natural species or medical treatments) can have partly different inherent or contextual characteristics, resulting in different bibliometric characteristics. This paper demonstrates that the newly developed 4D specialty approximation methodology has the ability to generate distinct, coherent results from seed directories in closely related specialties, reflecting known differences in publication and citation characteristics. This requirement being met, a next step is to investigate the approximation's adequacy (inclusion of peers, confirmation by scientists, potential bias, ways to enhance specialty coverage or delineation precision). A sufficient level of confidence reached, utility as a basis for assessment, trend analysis, recommendation, benchmarking, and distinction between different kinds of research (basic/applied, theoretical/empirical, mono-disciplinary/interdisciplinary) are among potential paths for exploration.